\documentclass[10pt,twocolumn]{article} 
\usepackage{simpleConference}
\usepackage{times}
\usepackage{graphicx}
\usepackage{amssymb}
\usepackage{url,hyperref}

\usepackage{natbib}
\usepackage{listings}
\usepackage{multirow}
\usepackage{tikz}
\usetikzlibrary{arrows,shapes,backgrounds}
\usepackage{verbatim}
\usepackage{array}
\usepackage{amsmath}
\usepackage{authblk}
\usepackage{placeins}
\usepackage{afterpage}
\usepackage{float}

\usepackage{caption}
\begin{document}
\tikzstyle{every picture}+=[remember picture]
\tikzstyle{na} = [baseline=-.5ex]

\title{Total Error Sheets for Datasets (TES-D) \\
\large A Critical Guide to Documenting Online Platform Datasets}

\author[1,2]{Leon Fröhling}
\author[1,2]{Indira Sen}
\author[1]{Felix Soldner}
\author[3]{Leonie Steinbrinker}
\author[1]{Maria Zens}
\author[1,4]{Katrin Weller}

\affil[1]{GESIS - Leibniz Institute for the Social Sciences, Cologne, Germany \authorcr
  \{\tt leon.froehling, indira.sen, felix.soldner, maria.zens, katrin.weller\}@gesis.org}
\affil[2]{RWTH Aachen University, Aachen, Germany}
\affil[3]{Leipzig University, Leipzig, Germany \authorcr
  \{\tt leonie.steinbrinker@uni-leipzig.de\}}
\affil[4]{CAIS - Center for Advanced Internet Studies, Bochum, Germany}

\maketitle
\thispagestyle{empty}

\begin{abstract}
This paper proposes a template for documenting datasets that have been collected from online platforms for research purposes. The template should help to critically reflect on data quality and increase transparency in research fields that make use of online platform data. The paper describes our motivation, outlines the procedure for developing a specific documentation template that we refer to as TES-D (Total Error Sheets for Datasets) and has the current version of the template, guiding questions and a manual attached as supplementary material. The TES-D approach builds upon prior work in designing error frameworks for data from online platforms, namely the Total Error Framework for digital traces of human behavior on online platforms (TED-On, \citet{sen2021total}).
\end{abstract}

\section{Introduction and Motivation}
Our work is motivated by current research practices in interdisciplinary fields like computational social science (CSS), internet research, web science and various individual disciplines that are using data collected from online platforms as a form of research data. Researchers are interested in the ``digital traces" that people leave when using online platforms as part of their everyday life. These traces result in digital behavioral data (DBD) that can be collected from the platforms (e.g., social media like YouTube, Twitter, Wikipedia, but also shopping portals, search engines, online maps and many more). While the variety of data types and the sizes of the datasets available from these platforms offer great opportunities for meaningful insights into opinions, communication and other forms of human behavior, both on the platforms and beyond, these novel types of data also come with various challenges that may impact aspects of data quality. Most fundamentally, the data are ``found" – in the sense that they are typically not designed for the research purpose that they are collected for – and are thus produced rather independently from a specific research context. Researchers must find their own ways for collecting data from platforms (which often appear as ``black boxes", opaque in their exact functioning), and the many design decisions made in collecting and processing these digital traces have a direct influence on the characteristics and quality of the resulting datasets. Since most researchers use their very own setups and pipelines for collecting and processing the data and since the online platform environments tend to change over time (new tools, new versions of access rights and restrictions), a need for careful inspection and documentation of the data curation processes and the resulting datasets arises. 

Data documentation practices are often connected to data archiving and are established in some research fields. Documentation may for example be based on controlled metadata (as for example provided by the Data Documentation Initiative, DDI).\footnote{https://ddialliance.org/} While there are no established, standardized documentation practices in computational social science, some related areas have already established useful standards. For example, in machine learning (ML) research, and natural language processing (NLP) in particular – disciplines that draw heavily on textual data collected from the web for many of their applications – a few solutions for the documentation of datasets have been implemented in recent years. Most of these efforts come from a small but growing community of Responsible AI researchers trying to raise awareness for issues of representation, fairness, transparency, and responsibility in a fast-paced research field. The idea behind approaches like Datasheets for Datasets \citep{gebru2021datasheets} or Data Statements for NLP \citep{bender2018data} is to transparently document and communicate the characteristics of datasets collected for the training and evaluation of machine learning models, thereby helping to expose potential biases and limitations. As the outputs and performances of ML models are directly dependent on the data they are trained on, any deployment of an ML model runs the danger of reproducing the biases and limitations inherent to the data, making the task of identifying and documenting them even more important. 

A different (and previously unrelated) approach for reflecting on data quality can be found in the social sciences, rooted in the extensive use of survey data. Here researchers established ``error frameworks" to identify potential sources of bias during the research process. Typically, error frameworks allow for the systematic assessment of the influence that research design choices have on the results of a study \citep{groves2010total}. One of the most prominent examples from survey research is the Total Survey Error framework by \citet{groves2011survey}. The Total Error Framework for Digital Traces of Human Behavior on Online Platforms (TED-On) takes this idea and transfers it to research with data collected from online platforms, thereby helping to expose challenges associated with the collection of traces, the underlying population that produced them, the meaning encoded in these traces, and the role of the platform in the trace generation process \citep{sen2021total}.  

Research with digital traces spans several disciplines that approach quality concepts of this data from various epistemological positions. Despite different disciplinary quality concepts, there is a growing consensus on the need for documentation to improve the quality of datasets obtained from digital traces \citep{olteanu2019social}. Current work in understanding and documenting the quality of this type of data has primarily approached it through either computational or quantitative lenses.  

We propose the TES-D as a hands-on tool that brings together the established data documentation practices from ML research and the idea of reflecting on research processes with the help of error frameworks as a contribution to the discourse on data quality for digital traces. By making its development process user-centered and interdisciplinary right from the start, the resulting documents should be accessible to various stakeholders and researchers interested in working with DBD. By integrating an error-focussed perspective on the data collection process with the idea of using datasheets to document the characteristics of datasets collected for research, we hope to both raise awareness for the challenges associated with DBD, and to facilitate the process of reflecting and documenting them. 

\section{Developing a Documentation Template for Online Platform Datasets}
Starting from the idea of combining the approaches of carefully examining and documenting the datasets used for research with error frameworks that help to structure and systematize the research processes during which these datasets are created, we first identified one documentation approach and one error framework to base our work on. 

For a documentation approach, we used Datasheets for Datasets, introduced by \citet{gebru2021datasheets}. We chose this approach as it is constructed around an abstracted research process that aligns well with the idealized process we have in mind when thinking about research with DBD. Along the different stages of this research process, the authors propose different questions that dataset creators are expected to answer in order to document their dataset. Even though their documentation approach was tailored for ML datasets, many of their questions transfer well to our purpose, as ML data and DBD are often quite similar, drawing heavily on innovative collection methods for many types of web data. 

We chose the TED-On \citep{sen2021total} as the backbone of our documentation approach since it is exactly targeted at working with DBD. It features an abstracted research process with clearly distinguished and well-defined steps, from construct definition and platform selection to data collection, data processing, and data analysis. For each of these steps, the TED-On introduces the design decisions that researchers face and discusses how they might impact the research results. Compared to similar efforts, the TED-On maintains the separation of errors into measurement and representation errors introduced in the Total Survey Error framework by \citet{groves2011survey}, describing whether they impact the measurement of a construct or the representativity of a study relative to some target population. 

To bring the systematic discussion of errors associated with the collection of DBD from the TED-On framework into a more concrete form, we selected those questions from Datasheets for Datasets that we deemed relevant for the documentation of DBD, mapping them onto the different steps of the research process identified in \citet{sen2021total}. While this already covered many of the most important aspects discussed in the TED-On, we made an additional effort to identify those aspects of the TED-On that were not already covered by the questions in \citet{gebru2021datasheets}. Furthermore, we added the section ``General Characteristics" to the TED-On categories to ensure that all relevant characteristics of a dataset – including (meta-)information like the creators of the dataset or distribution details – are covered. 

To make our TES-D approach operable, we designed different formats to interact with it, allowing researchers to integrate the systematic reflection on error sources and their documentation into their research workflows. Taking inspiration from the types of documentation sheets introduced by \citet{gebru2021datasheets} or \citet{bender2018data}, we first created a simple list of questions organized by categories. Retaining the schematic structure introduced through the TED-On framework, we also created a template version of the TES-D, which offers additional visual guidance for the documentation process. Lastly, we created a manual that provides additional information and context on the design decisions researchers face and discusses their potential to introduce different types of bias and errors into the data collection. We hope that this in-depth discussion of the research process makes it more accessible to researchers that might not be very familiar with the particularities of this type of data. 

The final phase of our development process consists of an iterative process of surveying various users of the TES-D, collecting their expectations and needs concerning the documentation of DBD and using these insights to improve the accessibility and usability of our materials. This user-centered development aims to improve interdisciplinarity and inclusivity, opening the TES-D approach to a wide range of potential users with various backgrounds and levels of expertise. By breaking down the often abstract and complex impacts of research design decisions in working with DBD into concrete and actionable questions, we aspire to facilitate a systematic and guided reflection on data collection for anyone interested in working with DBD.

\section{Provided Materials}
Based on existing documentation approaches and error frameworks, we propose TES-D in three different yet interconnected formats, each emphasizing separate aspects of the documentation process.\\

The different formats are: 
\begin{enumerate}
    \item A diagram-based template, showing the abstracted research process with its different steps and error components. The template is designed to better structure the documentation process, offering space for the user’s notes on the different errors identified in the TES-D (Appendix A).
    \item A set of questions, categorized along the different steps of the abstracted research process, that dataset creators using the template should answer to document their data collection (Appendix B). 
    \item A manual providing context on the different steps of the abstracted research process and on the relevant questions. Details provided in the manual include explanations and reasons why these questions are considered important for the documentation of data collection (Appendix C). 
\end{enumerate}

Additionally, we share an example of how the answers to the set of documentation questions could look like, documenting the \textit{Call me sexist, but..} dataset shared in \citet{samory2021call} (Appendix D).

\section{Next Steps}

We will leverage a human-centered approach through qualitative surveys conducted with participants before and after they are asked to apply the TES-D framework on a dataset of their choice. By surveying their opinions on data documentation processes and their experience of applying the TES-D framework specifically, we will better understand the needs of researchers and practitioners and the usability of the proposed documentation framework. Based on their feedback, responses, and our observations, we will incorporate improvements into the design of TES-D. Thus, allowing us to iteratively design contextualized documentation sheets for various use cases, data types, and stakeholders. Concurrently, we are working on an implementation of the TES-D approach into a popular programming environment. The idea is to lower the burden associated with data documentation by directly integrating the documentation process into the dataset creation workflow, facilitating the creation of dataset documentation sheets. This endeavor is supported by a SAGE Methods grant,\footnote{https://ocean.sagepub.com/blog/widd} and a first demo of the implementation is available via a GitHub repository.\footnote{https://github.com/gesiscss/widd-example}

\bibliographystyle{apalike}
\bibliography{main}

\appendix
\section*{Appendix}
\section{TES-D Template}


\begin{figure}[H]
\begin{center}
\onecolumn\includegraphics[width=0.8\textwidth,height=0.8\textheight]{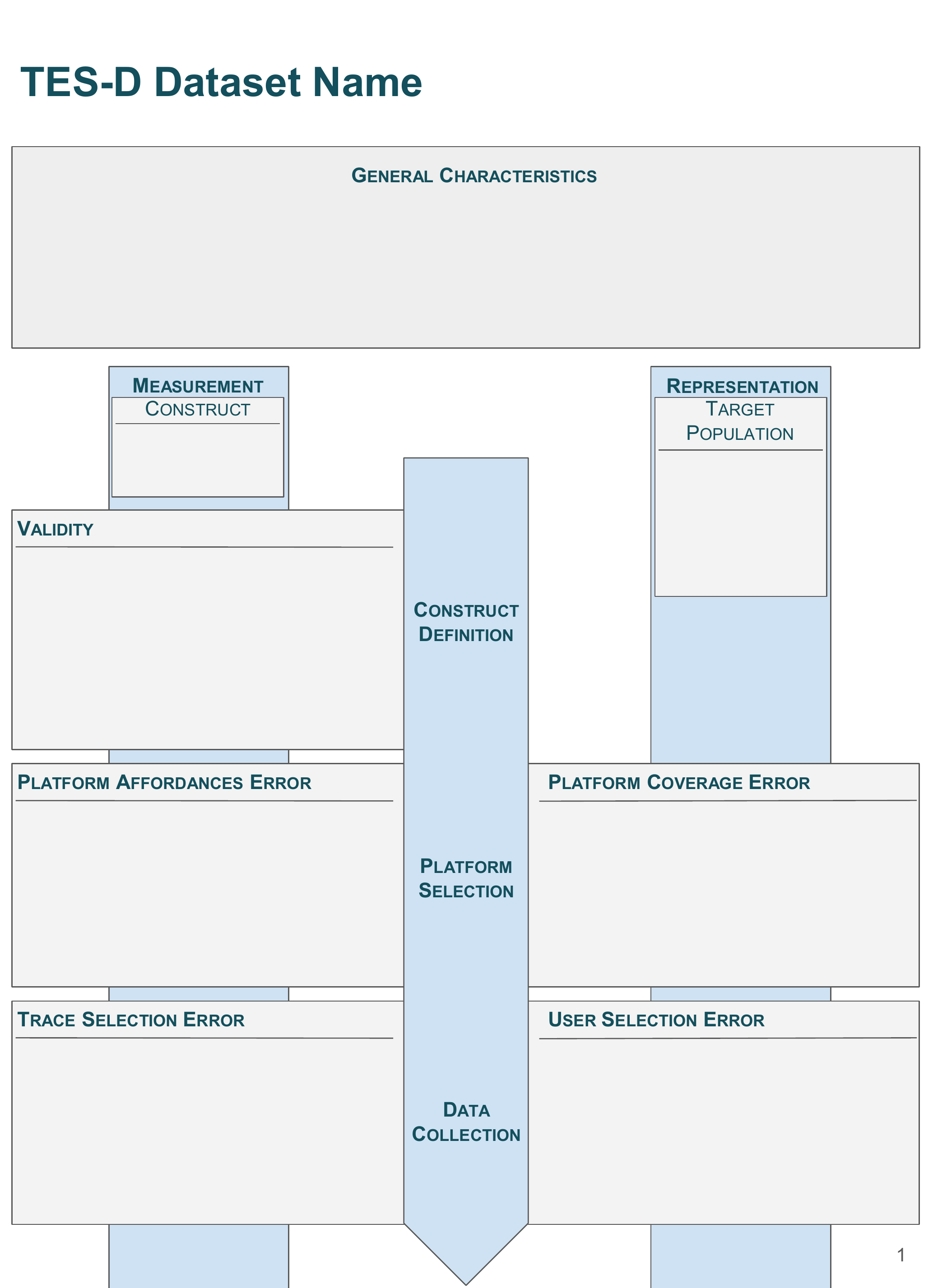}
\end{center}
\caption{First page of the diagram-based template.}
\label{fig:template1}
\end{figure}

\begin{figure}[H]
\begin{center}
\onecolumn\includegraphics[width=0.8\textwidth]{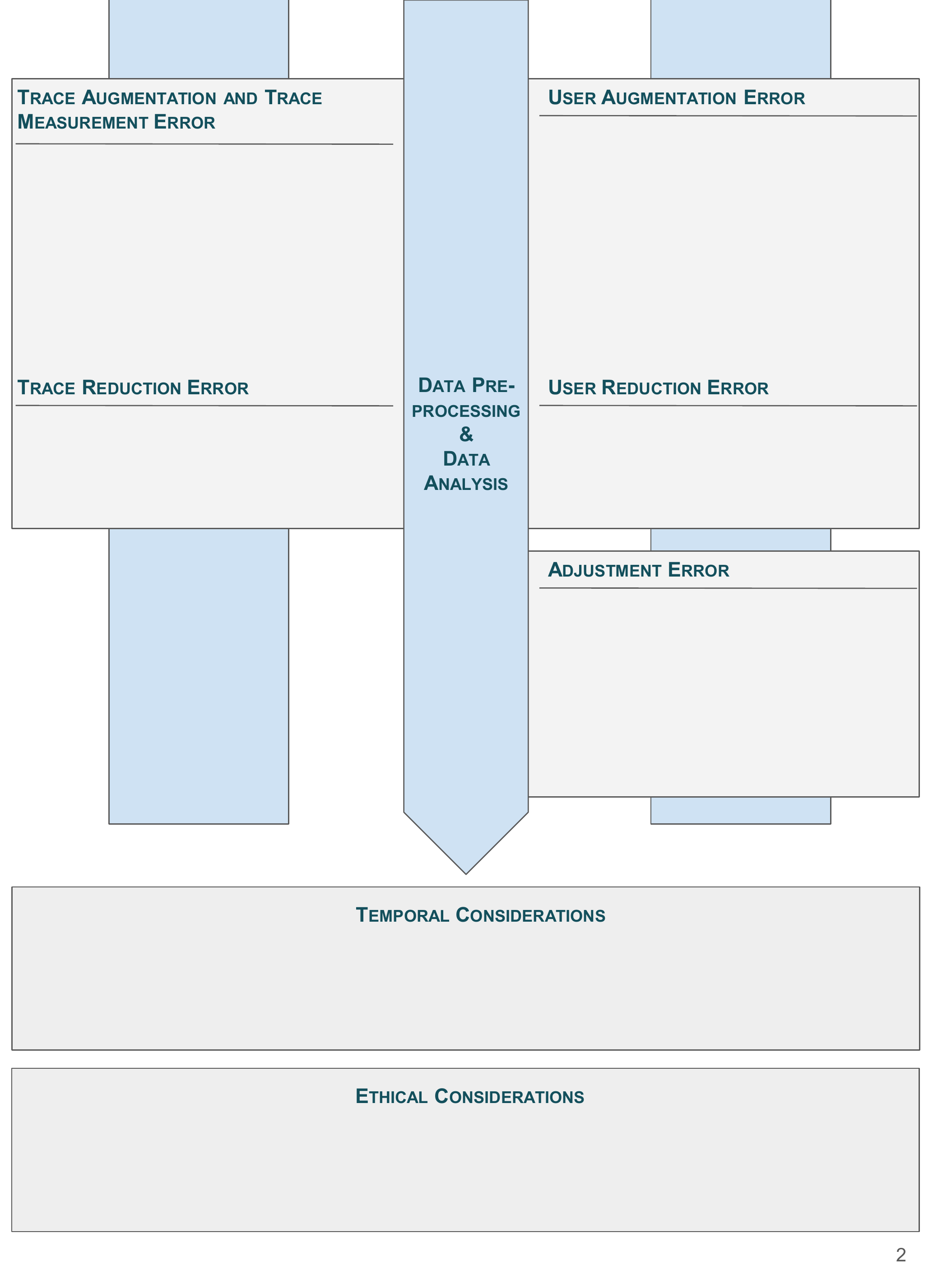}
\end{center}
\caption{Second page of the diagram-based template.}
\label{fig:template2}
\end{figure}

\clearpage

\section{TES-D Documentation Questions}

Please answer the following questions to document your dataset. More information on the different questions can be found in the corresponding manual. There is space to cite any additional materials you reference in your responses underneath the set of questions.\\

\begin{enumerate}
    \item \textbf{\underline{General Characteristics}} 
    \begin{enumerate}
        \item[1.1] Who collected the dataset, and who funded the process? 
        \item[1.2] Where is the dataset hosted? Is the dataset distributed under a copyright or license?
        \item[1.3] What do the instances that the dataset comprises represent? What data does each instance consist of?
        \item[1.4] How many instances are there in total in each category (as defined by the instances’ label), and – if applicable – in each recommended data split?
        \item[1.5] In which contexts and publications has the dataset been used already?
        \item[1.6] Are there alternative datasets that could be used to measure the same or similar constructs? Could they be a better fit? How do they differ?
        \item[1.7] Can the dataset collection be readily reproduced given the current data access, the general context, and other potentially interfering developments?
        \item[1.8] Were any ethics review processes conducted?
        \item[1.9] Did any ethical considerations limit the dataset creation? 
        \item[1.10] Are there any potential risks for individuals using the data? Does the data contain any disturbing images or texts? Could the content evoke psychological distress ?
    \end{enumerate}
    \item \textbf{\underline{Construct Definition}} \\
    \textbf{Validity}
    \begin{enumerate}
        \item[2.1] For the measurement of what construct was the dataset created?
        \item[2.2] How is the construct operationalized? Can the dataset fully grasp the construct? If not, which dimensions are left out? Have there been any attempts to evaluate the validity of the construct operationalization? 
        \item[2.3] What related constructs could (not) be measured through the dataset? What should be considered when measuring other constructs with the dataset? 
        \item[2.4] What is the target population? 
        \item[2.5] How does the dataset handle subpopulations?
    \end{enumerate}
    \item \textbf{\underline{Platform Selection}} \\
    \textbf{Platform Affordances Error}
    \begin{enumerate}
        \item[3.1] What are the key characteristics (relevant to the collected data) of the platform at the time of data collection?
        \item[3.2] What are the effects of the terms of service of the platform on the collected data?
        \item[3.3] What are the effects of the sociocultural norms of the platform on the collected data? 
        \item[3.4] How were the relevant traces collected from the platform? Are there any technical constraints to the data collection method? If so, how did those limit the dataset design?
        \item[3.5] In case multiple data sources were used, what errors might occur through their merger or combination?
    \end{enumerate}
    \textbf{Platform Coverage Error}
    \begin{enumerate}
        \item[3.6] What is known about the platform population? 
    \end{enumerate}
    \item \textbf{\underline{Data Collection}} \\
    \textbf{Trace Selection Error}
    \begin{enumerate}
        \item[4.1] How was the data associated with each instance acquired? On what basis were the trace selection criteria chosen? 
        \item[4.2] Was there any data that could not be adequately collected?
        \item[4.3] Is any information missing from individual instances? Could there be a systematic bias?
        \item[4.4] Does the dataset include sensitive or confidential information?
    \end{enumerate}
    \textbf{User Selection Error}
    \begin{enumerate}
        \item[4.5] Does the dataset contain all possible instances or is it a sample (not necessarily random) of instances from a larger set? If the dataset is a sample from a larger set, what was the sampling strategy?
        \item[4.6] What is known about the dataset population? Are there user groups systematically in- or excluded in/ from the dataset in direct consequence of the trace selection criteria?
        \item[4.7] Over what timeframe was the data collected, and how might that timeframe have affected the collected data?
        \item[4.8] If the dataset relates to people, how did they consent to collecting and using their data?
        \item[4.9] Does the data include information on minors? 
    \end{enumerate}
    \item \textbf{\underline{Data Preprocessing and Data Augmentation}} \\ 
    \textbf{Trace Augmentation and Trace Measurement Error}
    \begin{enumerate}
        \item[5.1] Is there a label or target associated with each instance? If so, how were the labels or targets generated? 
        \item[5.2] If automated methods were used, how does method performance impact the augmentations?
        \item[5.3] If human annotations were used, who were the annotators that created the labels? How were they recruited or chosen? How were they instructed?  How were they remunerated?
        \item[5.4] If the final label was derived from multiple annotations, how was this done?
        \item[5.5] Have there been any attempts to validate the labels?
        \item[5.6] How could the data be misused?
        \item[5.7] Could the dataset in any way contribute to the creation or reinforcement of social inequality? 
    \end{enumerate}
    \textbf{User Augmentation Error}
    \begin{enumerate}
        \item[5.8] Have attributes and characteristics of individuals been inferred?
        \item[5.9] Is it possible to identify individuals either directly or indirectly from the data?
    \end{enumerate}
    \textbf{Trace Reduction Error}
    \begin{enumerate}
        \item[5.10] Have traces been excluded? Why and by what criteria?
    \end{enumerate}
    \textbf{User Reduction Error}
    \begin{enumerate}
        \item[5.11] Have users been excluded? Why and by what criteria?
    \end{enumerate}
    \textbf{Adjustment Error}
    \begin{enumerate}
        \item[5.12] Does the dataset provide information to adjust the results to a target population? If so, is this information inferred or self-reported?
    \end{enumerate}
\end{enumerate}

\textbf{References}\\
Please use this space to cite any publications, websites or other sources you might have referenced in your responses.\\

\clearpage
\setcounter{footnote}{0}

\section{TES-D Manual}

This manual was designed to support the documentation of digital behavioral data(sets) collected from online platforms. Together with the TES-D template, this manual shall guide dataset creators through documenting the characteristics, limitations, and potentials of their data. By providing descriptions of the different steps necessary to collect digital behavioral data from online platforms as well as the various pitfalls and problems associated with it, this manual may also be consulted by researchers interested in working with this type of data to get a first idea of the challenges and difficulties they will likely face. 

The TES-D template and manual directly build on the TED-On framework,\footnote{Sen, I., Flöck, F., Weller, K., Weiß, B., and Wagner, C. (2021). A total error framework for digital traces of human behavior on online platforms. Public Opinion Quarterly, 85(S1):399–422. 4} an error framework for digital trace data collected from online platforms. It identifies several steps that make up the abstracted research process, and associates common design decisions and errors with them. The TED-On framework is inspired by the Total Survey Error framework from the social sciences, a framework that systematizes the different errors typical for survey research. It combines this systematic view on error and bias with the dataset documentation standards\footnote{Gebru, T., Morgenstern, J., Vecchione, B., Vaughan, J. W., Wallach, H., Daumé III, H., and Crawford, K. (2021). Datasheets for datasets. Communications of the ACM, 64(12):86–92} proposed and advocated by a community of researchers working in the field of responsible AI within the broader machine learning research landscape. The TES-D template and the chapters of this manual, are equivalent to the steps in the abstracted research process as identified by the TED-On framework. While steps like "Data Collection" and "Data Preprocessing" are part of any dataset creation process, other steps like "Data Analysis" might not be as intuitively seen as a part of a dataset documentation approach. However, since we consider the curation of a dataset in the context of a broader research process, we deem it important to document aspects that might influence subsequent analyses and uses of the dataset.  

Each chapter of the manual provides a brief description of the purpose of the step in creating the dataset and its influence on the research process. The chapters are then organized along the different potential errors identified in the TED-On framework. After a brief description of the error, the sub-chapters are structured by questions on the information necessary to identify and document systematic errors associated with digital behavioral data collected from online platforms. The questions are explained in short paragraphs, elaborating on the decisions researchers have to make during the collection process and the associated potential problems for the collected dataset. The framework aims to be comprehensive, not all sections and questions will apply to every dataset and may therefore be skipped. Questions relating to ethical considerations are not aggregated in a separate chapter but are placed among the questions of the step in the research process to which they thematically relate, emphasizing the need to critically reflect upon the ethical implications of the research design at every step of the process. \\

\textbf{\underline{1. General Characteristics}} \\
This section should provide the reader with an overview of the dataset, including its contents, access details, and the motivation for its creation. While the general characteristics do not contribute to the error-focus of the rest of the TES-D approach, they are necessary to make the documentation comprehensive and independent of additional resources. \\

\textit{1.1 Who collected the dataset, and who funded the process?} Please provide information on author affiliations, on contributors roles during the dataset creation, and correspondence addresses. Disclose how and by whom the dataset collection was funded. \\

\textit{1.2 Where is the dataset hosted? Is the dataset distributed under a copyright or license?} Provide the DOI of the dataset or a link to its (permanent) hosting location and access and licensing information. \\

\textit{1.3 What do the instances that the dataset comprises represent? What data does each instance consist of?} Does the dataset include different types of instances (e.g., users and their relations)? Can instances be interpreted as ``humans"? Is the data ``raw" (e.g., unprocessed text), or has it been pre-processed (e.g., features extracted from ``raw" data) or augmented with additional ``labels"? \\

\textit{1.4 How many instances are there in total in each category (as defined by the instances’ label), and – if applicable – in each recommended data split?} Provide a first impression of the distribution of the dataset by giving the number of instances per label category and data split. \\

\textit{1.5 In which contexts and publications has the dataset been used already?} If known to you, please provide information on the uses of the dataset. \\

\textit{1.6 Are there alternative datasets that could be used to measure the same or similar constructs? Could they be a better fit? How do they differ?} If you came across similar datasets during the creation process, please provide details here. Explain why you decided to collect the dataset rather than re-using an available alternative. \\

\textit{1.7 Can the dataset collection be readily reproduced given the current data access, the general context, and other potentially interfering developments?} Report if recent developments (e.g., discontinued APIs, change of terms of service) prevent the dataset from being readily reproduced. \\

\textit{1.8 Were any ethics review processes conducted?} Who reviewed the ethical considerations of the data collection and approved it? Did any difficulties arise during the ethics review process, and how were they addressed? \\

\textit{1.9 Did any ethical considerations limit the dataset creation?} In case you decided against the inclusion of certain instances or information based on ethical concerns, please provide information on your considerations here. \\

\textit{1.10 Are there any potential risks for individuals using the data? Does the data contain any disturbing images or texts? Could the content evoke psychological distress?} This applies to, e.g., any illicit, crime, or violence-related content. \\

\textbf{\underline{2. Construct Definition}} \\
The construct is the essence of what data-based research is interested in measuring. Without a clear and precise definition of the construct, any potential operationalization based on the available data will suffer from a lack of validity. If the dataset was not collected with a specific construct or purpose in mind, many questions in this section might not apply and may be skipped. However, we would still encourage readers to familiarize themselves with the important issues around the definition and operationalization of constructs raised in this section. Datasets are always part of a broader research landscape and might, if not by the original creators of the dataset, still be used for measuring a specific construct at a later stage. \\

\textbf{Validity} \\

Validity is concerned with demonstrating that measures obtained for a specific construct are both meaningful and useful. The careful design of a suitable construct is especially crucial before using digital behavioral datasets, as the connection of digital traces to concepts relevant to research must often be established first. \\

\textit{2.1 For the measurement of what construct was the dataset created?} Provide a concise and precise definition of the construct, ideally based on a sound theoretical foundation. Constructs often encompass a multitude of dimensions, requiring careful reflection and consideration on the questions of which aspects are most important for the research purpose and how they might best be measured. The resulting definition of the construct should inform all the following data collection procedures. \\

\textit{2.2 How is the construct operationalized? Can the dataset fully grasp the construct? If not, which dimensions are left out? Have there been any attempts to evaluate the validity of the construct operationalization?} Report how the construct definition is operationalized. The operationalization should link the data and the theoretical construct, capturing it in all its defined dimensions. Previous datasets covering the construct may help finding a suitable operationalization. An operationalization that (partially) measures a different construct is potentially problematic. In some cases, an insufficient operationalization cannot be avoided because of ethical reasons or the inherent complexity of the construct, which should be reported transparently. \\

\textit{2.3 What related constructs could (not) be measured through the dataset? What should be considered when measuring other constructs with the dataset?} Share your expertise on the topical domain of the dataset by pointing to relevant related resources and alternatives to the collected dataset. Such information will support  other researchers in making informed decisions about potentially re-using the collected dataset. \\

\textit{2.4 What is the target population?} The results of research building on a dataset will always be immediately valid only for the population of the dataset. If extrapolation to a target population outside of the dataset is desired, this target population must first be carefully defined. If the dataset has been collected with a concrete research question in mind, the population for which the dataset was collected should be defined here. \\

\textit{2.5 How does the dataset handle subpopulations?} If the dataset is on individuals and their characteristics, have there been attempts to ensure a balanced composition of the dataset? Does the dataset contain a decent amount of instances for all relevant subpopulations? \\

\textbf{\underline{3. Platform Selection}}\\
The selection of a platform determines what traces can be collected by the dataset creator(s). The platform’s available traces should align with the construct definition and operationalization. The ways in which the characteristics of the platform shape and influence the user behavior and, thereby, the traces in the dataset should be examined and documented.\\

\textbf{Platform Affordances Error} \\

The platform architecture affects the collected data. Researchers should reflect on the differences between various platforms and how platform characteristics and sociocultural norms shape and influence user behavior. By understanding the intricacies of the platform, dataset creators can better differentiate between the relevant, platform-independent behavior and platform-driven behavior. \\

\textit{3.1 What are the key characteristics (relevant to the collected data) of the platform at the time of data collection?} Naming and reflecting on the key characteristics of a platform helps to understand the type and format of the collected traces. \\

\textit{3.2 What are the effects of the terms of service of the platform on the collected data?} Moderation practices and ToS constitute the observable framework that shapes and delimits the communication and interaction processes of the platform population. They directly influence how content is moderated and what users post and share. \\

\textit{3.3 What are the effects of the sociocultural norms of the platform on the collected data?} On top of the observable framework imposed by the platform operators through the terms of service and the moderation practices, the user demographics and shared norms and values, are population-inherent factors that influence the expected and accepted behavior on the platform. Social desirability and perceptions of the audience and their expectations affect what users feel comfortable to share and with whom they are willing to interact publicly. \\

\textit{3.4 How were the relevant traces collected from the platform? Are there any technical constraints to the data collection method? If so, how did those limit the dataset design?} When data is collected through the official APIs provided by the platforms, the data collection is influenced by the type of data available, restrictions like rate limits that limit the volume of data that can be collected at once, or platform-sided sampling. The frequently changing permissions and limitations regulating the data access and the particularities of the data available through the APIs should be documented at the time of the data collection to allow traceability and comparability. Alternatively, individualized web scrapers may be used for data collection, which are also subject to potential restrictions such as limited speed, the handling of recurrent website changes, or area-restricted access as determined through IPs. The dataset creators should therefore specify the method used and document any associated restrictions and resulting problems, as well as their effects on the compilation and composition of the dataset. \\

\textit{3.5 In case multiple data sources were used, what errors might occur through their merger or combination?} The previously mentioned dataset characteristics of each data source should be compared and their compatibility tested before the data is merged or combined. How digital behavioral data might be used to enrich data from more traditional sources and how different datasets can be meaningfully combined are questions of ongoing research. \\

\textbf{Platform Coverage Error} \\
The particular characteristics of different platforms not only influence their user behavior but also attract different types of users. Additionally, the way that users interact on different platforms and the general internet access across different regions and socio-demographic or socio-economic groups change over time, influencing whose traces can be found on the platforms. \\

\textit{3.6 What is known about the platform population?} To validly infer from the population represented in a dataset collected from a specific platform to a target population outside of that particular platform, a good understanding of who is included in the dataset needs to be attained. User demographics and the temporal and geographical boundaries of the dataset should therefore be carefully documented. \\

\textbf{\underline{4. Data Collection}} \\
Within the constraints of data availability and technical limitations of the access offered by platforms, dataset collectors need to decide how to select traces and user information relevant for their study. The dataset creator has to calibrate the collection process so that all traces that contribute to the measurement of the construct and all users that are part of the target population are collected, without including traces and users that are irrelevant to the construct and target population. \\

\textbf{Trace Selection Error} \\
Depending on the chosen operationalization of the construct of interest, relevant traces for its calculation need to be collected. When collecting these traces, every detail of the collection process potentially influences the composition of the resulting dataset. However, not only the deliberate decisions of dataset collectors are relevant, but also the amount and type of traces made available from the platform. While the traces available for collection often present only a subset of the total traces in existence, in most circumstances, not all available traces can be collected, stored, and processed. \\

\textit{4.1 How was the data associated with each instance acquired? On what basis were the trace selection criteria chosen?} Collectors of datasets often rely on keyword-based queries to select only those traces relevant to the construct of interest. This involves identifying which traces are relevant, analyzing their characteristics, and formulating queries that combine keywords, filters, and logical operators so that relevant traces are collected, and irrelevant traces are avoided. To ensure that the collected data is adequate for the intended purpose, the assumptions justifying the used query should be made transparent. The approach of using lists of heuristic keywords to collect all data on a specific topic runs the risk of missing parts of the relevant data (``false negatives") due to the omission of meaningful keywords. In addition to missing parts of the relevant data, additional or alternative meanings of keywords can also lead to the inclusion of traces irrelevant to the data collection (``false positives"). Thus, the trace selection criteria and the considerations that led to their use should be documented, to transparently document the potential issues of including irrelevant and omitting relevant traces. \\

\textit{4.2 Was there any data that could not be adequately collected?} Ideally, only the deliberate query design would affect the data collection process. However, in practice, many other factors can lead to errors, noise or redundancies during data collection. Frequent issues are technical errors and disconnects that may lead to the loss of whole subsets of the data. Furthermore, users' privacy settings could systematically hinder the collection of certain traces from a very specific group of users. API usage limits might require workarounds during data collection that increase the likelihood of unintendedly including duplicate entries. \\

\textit{4.3 Is any information missing from individual instances? Could there be a systematic bias?} If individuals do not find themselves represented in the classification schemes utilized in the data, they might not (be able to) provide information on certain attributes (e.g., binary gender labels). Such issues can lead to the systematic omission of these individuals. \\

\textit{4.4 Does the dataset include sensitive or confidential information?} Such information may refer to individuals' religious beliefs, sexual orientations, or specifics about health. Are any data protection measures in place? What regulatory policies does the dataset adhere to (e.g., GDPR)?  \\

\textbf{User Selection Error}\\
While the Trace Selection Error is concerned with the systematic in- and exclusion of traces, the User Selection Error looks at users that are systematically omitted or included. It often builds directly on Trace Selection Error, especially if users are not explicitly collected but included as a consequence of the collected traces. \\

\textit{4.5 Does the dataset contain all possible instances or is it a sample (not necessarily random) of instances from a larger set? If the dataset is a sample from a larger set, what was the sampling strategy?} In contrast to the continuous stream of content generated on online platforms, most modes of data collection are static and thus (at least) limited by time, such that datasets will almost inevitably only cover parts of the platform under study. Apart from the temporal aspect, constraints like the data access granted via APIs or the processing and storage capacities of the collection process render it impossible to collect all available traces. The decisions that influence the amount and characteristics of the data that is ultimately collected should be made transparent so that their impact on the composition of the collected dataset can retrospectively be assessed. \\

\textit{4.6 What is known about the dataset population? Are there user groups systematically in- or excluded in/ from the dataset in direct consequence of the trace selection criteria?} After making any sampling procedures that occurred during data collection transparent, the final composition of the population covered in the dataset should be investigated. If the data is about individuals, relevant dimensions to be investigated for a balanced inclusion could be race, sex, and age. If certain groups are missing from the dataset, possible reasons should be investigated, to see whether they are excluded for one of the reasons described above or whether they are not relevant for the dataset and thus deliberately missed. The dataset composition, and especially any systematic in- or exclusions, should be documented and communicated transparently. \\

\textit{4.7 Over what timeframe was the data collected, and how might that timeframe have affected the collected data?} The role and importance of online platforms change over time, with new platforms attracting users from more established alternatives. To understand what population has been present on the platform during data collection, the time over which the data was collected, and a brief description of any relevant events that might have influenced the platform’s user base since then should be provided. \\

\textit{4.8 If the dataset relates to people, how did they consent to collecting and using their data?} Specify how individuals consented to the use of their data for research purposes and what possible restrictions apply. Describe how the given consent may be revoked or altered and how individuals that are part of the dataset may request the removal of their data. Describe how the collected data is updated to comply with the decisions of individuals or platforms to remove data from the public. \\

\textit{4.9 Does the data include information on minors?} If yes, specify how they are protected and how their guardians consented. \\

\textbf{\underline{5. Data Preprocessing and Data Analysis}} \\
As introduced in the TED-On framework, it is not always possible to completely disentangle the pre-processing and analysis steps undertaken to arrive at the final representation of the data. The TED-On framework differentiates these annotation efforts into the steps ``Trace Augmentation" and ``Trace Measurement", associated with the steps ``Data Preprocessing" and ``Data Analysis" of the abstracted research process. While ``Trace Augmentation" and ``Trace Measurement" might have the same goal of enriching the data with information (like additional labels or alternative representations) and rely on the exact same methods and models to achieve this, their position in and significance for the process of creating a dataset are different. In extracting the relevant construct from the raw data, ``Trace Measurement" is always the last step, and often builds on features and representations extracted through previous ``Trace Augmentation" steps. \\

\textbf{Trace Augmentation and Trace Measurement Error} \\
Datasets that not only capture instances and their characteristics but also augment those mere observations with associated labels representing some form of additional knowledge on these instances, are a valuable resource in CSS research. While the original idea of these datasets has been to collect reliably true values only, it is increasingly common to approximate these true values via crowd workers or computational models explicitly trained for the task, introducing additional uncertainty regarding data quality. \\

\textit{5.1 Is there a label or target associated with each instance? If so, how were the labels or targets generated?} Describe what parts of the dataset have been generated via automated methods or human annotation, and indicate which methods were used. Explain why these methods have been chosen over available alternatives. \\

\textit{5.2 If automated methods were used, how does method performance impact the augmentations?} In using ML models to extract additional features from raw data, the performance often decreases significantly for input data that is different from the data that the model was trained on (e.g., from a different source, regarding a different topic, in a different conversational context). If the impact of the applied model on the annotations cannot be directly assessed and measured, the steps taken and methods used should be documented so that they might be subject to future investigation. \\

\textit{5.3 If human annotations were used, who were the annotators that created the labels? How were they recruited or chosen? How were they instructed?  How were they remunerated?} Cultural communities define themselves by their common knowledge of concepts and their relationships. Therefore human annotations will never represent a universal perspective and a globally true value, but it is rather the annotators’ unique demographics that will shape and influence any human annotated dataset. Thus, the annotation process should be documented, including whose perspectives might have manifested in the dataset, as well as what efforts have been made to recruit and instruct a diverse and qualified set of annotators. Additionally, information on the incentives and remunerations should be reported.\\

\textit{5.4 If the final label was derived from multiple annotations, how was this done?} If different annotations are aggregated to produce a final label, seen frequently in attempts to reduce the influence and bias of individual annotators, the mechanism by which the final label was extracted should be explained. If annotations on different dimensions and aspects of the construct are aggregated, or different scales are combined to produce the final label, the coding scheme should be provided. \\

\textit{5.5 Have there been any attempts to validate the labels?} Describe whether the final labels of the dataset were validated and that they measure what they are supposed to measure, or, whether they suffer from biases introduced by using ML models or through human annotators. Validation efforts could encompass manual inspection by the dataset creators, the calculation of inter-annotator correlation measures, or the aggregation and comparison to established knowledge. \\

\textit{5.6 How could the data be misused?} Could serious societal harms or dangers result from intentional misuse of the dataset? Could the dataset unintentionally expose individuals or social groups or lead to abuse or repercussions by revealing sensitive information? If so, is it responsible to build the dataset in the first place? \\

\textit{5.7 Could the dataset in any way contribute to the creation or reinforcement of social inequality?} Reflect upon the potential of research or applications built on top of the dataset to reinforce social inequality. For example, this could be research that exposes certain characteristics of subpopulations that might subsequently experience discrimination based on these characteristics or applications that do not work (reliably) for some subpopulations, excluding them from the use of these applications. \\

\textbf{User Augmentation Error} \\
Characteristics and attributes of individuals that could not be collected might also be inferred from the available data. Automated methods and manual labeling are frequently used for the user augmentation task, but both can lead to the misrepresentation of individuals. \\ 

\textit{5.8 Have attributes and characteristics of individuals been inferred?} Inferring attributes from individuals always implies their reduction into the categories chosen by the researchers. If individuals do not fit into any of those categories, the categorization risks losing nuance and misrepresenting individuals, and, in the worst case, denies them their existence outright. \\

\textit{5.9 Is it possible to identify individuals either directly or indirectly from the data?} Specify whether there have been attempts to pseudonymize the data by removing personal identifiable information. However, some types of data might contain highly detailed information about individuals, possibly allowing for their identification. \\

\textbf{Trace Reduction Error} \\
To avoid the inclusion of traces irrelevant to the purpose of the data collection, filters and heuristics may be used to remove such traces from the dataset. Irrelevant traces could be traces identified as spam, traces in a language or format not covered by the purpose of data collection, or traces that have mistakenly been included for one of the reasons described in the Trace Selection Error section and are therefore filtered out. \\

\textit{5.10 Have traces been excluded? Why and by what criteria?} Describe whether and with which methods traces irrelevant for the measurement have been filtered from the dataset. Why were they deemed irrelevant? Similar to the over- and underinclusion described in the Trace Selection Error section, heuristics and filters to remove irrelevant traces might systematically drop relevant traces or include irrelevant ones. \\

\textbf{User Reduction Error} \\
To avoid the inclusion of users irrelevant to the purpose of the data collection, filters and heuristics may be used to remove such users from the dataset. Frequently, users associated with inauthentic behavior (e.g., trolling, automation) are excluded from the dataset, as their behavior is not considered authentically human. \\

\textit{5.11 Have users been excluded? Why and by what criteria?} Similar to reducing traces, removing certain types of users from the dataset might be necessary. Describe what type of user is considered irrelevant for data collection, and explain why. How are these users filtered out? \\

\textbf{Adjustment Error} \\

To make up for the non-probabilistic sampling inherent to platform-based data collections, researchers might apply reweighting techniques to adjust results to the target population. \\

\textit{5.12 Does the dataset provide information to adjust the results to a target population? If so, is this information inferred or self-reported?} Reweighting techniques work by aligning the demographics of the platform population represented in the dataset with the demographics of the target population. For this adjustment to be successful, demographic information must be included in the dataset. Since this information is often not available from online platforms, the process of acquiring these traces as well as their quality should be reported in great detail. \\

\clearpage
\setcounter{footnote}{0}

\section{TES-D Example \textit{Call me sexist, but..}}

This TES-D serves to document the dataset presented in Samory et al. (2021). The following documentation was created by Leon Fröhling (\texttt{leon.froehling@gesis.org}), Indira Sen, and Felix Soldner. \\

\textbf{\underline{1. General Characteristics}}\\

\textit{1.1 Who collected the dataset, and who funded the process?}\\
The dataset was collected by a team of researchers associated with the GESIS - Leibniz Institute for the Social Sciences CSS department. In particular: Samory, Mattia (Head Author); Sen, Indira (Data Curator); Kohne, Julian (Data Curator); Floeck, Fabian (Project Lead); Wagner, Claudia (Project Lead).\\

\textit{1.2 Where is the dataset hosted? Is the dataset distributed under a copyright or license?}\\
The dataset is hosted at the GESIS datorium (DOI: https://doi.org/10.7802/2251), and can be freely accessed after registration. It is distributed under the CC BY-NC-SA 4.0 license.\\

\textit{1.3 What do the instances that comprise the dataset represent? What data does each instance consist of?}\\
The instances in the dataset represent scale items, tweets, or adversarial augmentations created by crowdworkers. The instances have the following data fields: a unique ID, the dataset they are coming from, a toxicity score from the Perspective API\footnote{https://www.perspectiveapi.com/}, a binary sexism classification, and a second ID if the instance is a modification of another instance. A supplementary dataset, providing the individual labels for sexism due to both content and phrasing, as well as the annotations from each crowdworker that, in aggregate, inform the final sexism classification, is also available. \\ 

\textit{1.4 How many instances are there in total in each category (as defined by the instances’ label), and - if applicable - in each recommended data split?}\\
There are a total of 13,631 (1,809 labeled as sexist) instances in the dataset. Of these, 2,292 instances are adversarial examples. 
There are different sources for the instances, with the following numbers of instances coming from each of the sources: 
\begin{itemize}
    \item 1,080 (189) from the benevolent dataset, 402 of which are adversarial augmentations 
    \item 2,431 (790) from the callme dataset, 1,151 of which are adversarial augmentations 
    \item 1,257 (290) from the hostile dataset, 579 of which are adversarial augmentations 
    \item 878 (540) from the scales dataset, 135 of which are adversarial augmentations 
    \item 7,985 (0) from other datasets, 25 of which are adversarial augmentations
\end{itemize}

\textit{1.5 In which contexts and publications has the dataset been used already?}\\
The dataset has been used by Samory et al.(2021) to evaluate the reliability of machine learning models trained for the task of detecting sexism in tweets. It has also been used in follow-up work by Sen et al. (2022) for assessing the efficacy of adversarial augmentation.\\

\textit{1.6 Are there alternative datasets that could be used for the measurement of the same or similar constructs? Could they be a better fit? How do they differ?}\\
While there are many datasets available for different aspects of sexism on social media platforms, this dataset proposes a method to map out the various aspects of sexism as a construct as comprehensively as possible, allowing to operationalize and measure the construct of sexism as nuanced as possible. Depending on the nuance and dimension of sexism a researcher is interested in, any of these specific datasets might be an appropriate fit, while this dataset works as a more comprehensive and general one. Other alternative datasets could be related to abusive language in general as well as more specific datasets on misogyny.\\ 

\textit{1.7 Can the dataset collection be readily reproduced given the current data access, the general context and other potentially interfering developments?}\\
While the scales used for the dataset should still be available as part of scientific publications (subject to license agreements), the authors release the redacted tweet texts rather than the IDs for safeguarding data subjects' privacy, which means that the dataset can be directly reused for future work. However, to replicate the data collection from scratch, the tweet IDs would be required.


\textit{1.8 Were any ethics review processes conducted?}\\
There was no IRB or ethics review process conducted. The work does not fall under research with human subjects and care was taken to ensure that crowdworkers were compensated in accordance with fair pay guidelines and debriefed about the potentially harmful nature of the content they would be annotating.\\
 
\textit{1.9 Did any ethical considerations limit the dataset creation?}\\
For ethical reasons, the adversarial augmentation was limited to turning sexist instances into non-sexist instances, and not vice-versa.\\

\textit{1.10 Are there any potential risks for individuals using the data? Does the data contain any disturbing images or texts? Could the content evoke psychological distress?}\\
Due to the nature of the dataset, i.e., manifestations of explicit and nuanced types of sexism, it can be potentially upsetting for data stewards.\\

\textbf{\underline{2. Construct Definition}}

\textbf{Validity} 

\textit{2.1 For the measurement of what construct was the dataset created?}\\
The dataset was created to measure the construct of sexism in a more comprehensive manner compared to state-of-the-art research, which is heavily focused on overt forms of sexism, with the goal of improving sexism detection online. The dataset creators provide detailed information on the literature and scales covered by their construct definition in Samory et al. (2021). 
Based on the scales collected from the literature, four new categories on sexism are iteratively formed: behavioral expectations, stereotypes and comparisons, endorsements of inequality, denial of inequality and rejection of feminism. Additionally, the dataset creators not only consider the contents of a text as potentially sexist, but also its phrasing. \\  

\textit{2.2 How is the construct operationalized? Can the dataset fully grasp the construct? If not, which dimensions are left out? Have there been any attempts to evaluate the validity of the construct operationalization?}\\
The construct is operationalized through a coding scheme which is then used by annotators hired for the task of determining whether an instance is sexist or not. Those crowdworkers follow the instructions laid out in a codebook, available from Samory et al. (2021). The coding scheme was validated by asking five crowdworkers to apply the coding scheme to the ground truth data. The resulting annotations’ majority verdict (min. three out of five) corresponded with the ground truth label in 86\% of the cases. Further information on the measures taken to ensure qualitatively good and valid annotations are reported in response to questions \textit{5.3}. 
Even though the dataset creators made an attempt to condense all the collected aspects and dimensions of sexism into their final categories, it is unlikely that textual content on social media covers all possible manifestations of sexism. As an example, sexism reflected in patronizing behavior towards women cannot be covered in this dataset. 
\\ 

\textit{2.3 What related constructs could (not) be measured through the dataset? What should be considered when measuring other constructs with the dataset?}\\
Scales referring to constructs ``similar to sexism" are also included in the dataset. This could lead to issues related to convergent validity, if aspects outside of a specific definition of sexism are included in the dataset. However, for the same reason, the dataset could also be applicable to constructs closely related to sexism.\\  

\textit{2.4 What is the target population?}\\
The authors do not explicitly define a target population for their study, but mention that their codebook should apply to ``sexism in social media". The target population arising from this would thus be the population of all users of social media, or, if considering that the data collection is limited to English tweets, the population of English-speaking Twitter users.\\

\textit{2.5 How does the dataset handle subpopulations?}\\
The dataset does not explicitly include humans as subjects nor include demographic information of the creators of the different instances. It therefore does not explicitly identify, define, or act upon subpopulations.\\

\textbf{\underline{3. Platform Selection}}\\

\textbf{Platform Affordances Error}\\

\textit{3.1 What are the key characteristics (relevant to the collected data) of the platform at the time of data collection?}\\
Key characteristics of Twitter at the time of data collection include the 280 character limit for tweets.

\textit{3.2 What are the effects of the terms of service of the platform on the collected data?}\\
As the terms of service (ToS) at the time of data collection have not been documented, the following is based on the current version.
The ToS state that the user is responsible for any content they provide, as well as for their compliance with applicable laws, rules, and regulations. It furthermore identifies severe, repetitive usage of [...] sexist tropes where the primary intent is to harass or intimidate others as a reason for a tweet’s removal. Through these rules and their enforcement through moderation practices, Twitter tries to remove sexist content from the platform. 
Therefore, the efficiency and efficacy of the moderation practices influences how many sexist tweets are available for collection from the platform at any point in time, and the design of the moderation practices determines what types of sexist content are removed and what types remain on the platform. Changes in the ToS and changes in their enforcement through moderation practices would thus be reflected in the collected data.\\ 

\textit{3.3 What are the effects of the sociocultural norms of the platform on the collected data?}\\
At the time of data collection, Twitter was well established as the most popular platform for certain types of users to comment on acute events and issues of general importance for the society. The debates on Twitter are often perceived as polarized and certain topics tend to ``blow-up" on Twitter, being disputed about at great length and with great fervor by users. This has led to instances of trolling, where users would try to trigger such (exaggerated) reactions from other users by posting provocative, polarizing, or even straight-out abusive and harmful statements, including sexist tropes, even if with an implied distance or irony to it, or in some sort of meme-format, as with the phrase ``Call me sexist, but.." . The dataset is directly picking up on these patterns that are the result of the culture on Twitter.\\ 

\textit{3.4 How were the relevant traces collected from the platform? Are there any technical constraints to the data collection method? If so, how did those limit the dataset design?}\\
The tweets were collected via the Twitter streaming endpoint of the Twitter API. Tweets with the phrase ``call me sexist, but", posted between 2008 and 2019, were collected from the API.
Collecting data from the historic search endpoint means that, in theory, every tweet since the creation of Twitter back in March 2006 can be searched and thus potentially be included in the dataset. However, retrospective collection of tweets, as is always the case when using the historic search endpoint, does not allow for the collection of tweets that have been removed by Twitter, by the user, or that have been made private. Twitter might remove tweets that directly violate their rules, and also automatically removes tweets that were posted by users that have been (permanently) removed from the platform, e.g. for their failure to comply with the rules. A tweet would also not be accessible anymore if its author decided to delete it entirely, or if that author changed their account settings to private, thus preventing the tweet from being found by others.  
For this dataset, this could mean that not all tweets that were ever posted containing the phrase call me sexist, but are included in the dataset. Since the phrase is chosen for being and indicator of potentially sexist content that follows, it is rather likely that some of the tweets containing that phrase have either been deleted by the user for the backlash that the tweet received, or for further reflections that made the author uncomfortable with being associated with the tweet’s contents. The sexist nature of these tweets also makes it more likely that Twitter removes them due to content moderation, either by directly removing the tweet or indirectly, by suspending its author from the platform.\\

\textit{3.5 In case multiple data sources were used, what errors might occur through their merger or combination?}\\
The final dataset is a combination of different existing datasets for sexism on social media, a collection of psychological scale items used to measure sexism, as well as a newly collected Twitter dataset. 
Challenges of combining these different data sources include the mixture of different types of texts (i.e. scale items and tweets), which very likely have different characteristics, e.g. in terms of their length, their structure, their linguistic style, or their vocabulary. To a certain extent, this also holds true for the different datasets collected from Twitter, where differences in the collection strategy, the time of collection, or the pre-processing of the raw tweets might lead to artifacts in the data. 
However, by using a codebook specifically derived for this dataset and by re-annotating the entire dataset, a lot of effort is made to ensure the compatibility of instances coming from different sources.\\

\textbf{Platform Coverage Error}\\

\textit{3.6 What is known about the platform population?}\\
The Pew Research Center regularly conducts surveys to characterize the Twitter platform population (Wojcik and Hughes, 2019). They do so by asking their survey participants on their Twitter usage, as well as by asking for consent to analyze their Twitter data. Their findings are then reweighted to match the target population, which is Twitter users age 18 and older, living in the U.S.. 
For these adult Twitter users, it is found that media personalities, politicians and the public turn to social networks for real-time information and reactions to the day’s events. Twitter users, however, tend to be younger, are more likely to identify as Democrats, are better educated and have higher incomes than the overall population of U.S. adults. In themselves, Twitter users split into those that are very active, with the 10\% most active users being responsible for 80\% of all tweets in the U.S., and those that are inactive or turn to the platform only to keep up to date and to read, not to actively participate and engage. 
While there are many anecdotal insights into the demographics and motivations of Twitter users, only few actually reliable empirical analyses of the platform population exists, as many Twitter users use the platform anonymously or pseudonymously, or simply do not provide the relevant demographic information. Researchers then need to resort to inferring those missing demographics, which in many cases might be of questionable reliability (Buolamwini and Gebru, 2018).\\

\textbf{\underline{4. Data Collection}}\\

\textbf{Trace Selection Error}\\

\textit{4.1 How was the data associated with each instance acquired? On what basis were the trace selection criteria chosen?}\\
The instances collected for this dataset are the tweets collected through the historic search API containing the keyphrase ``call me sexist, but". The rationale behind this choice of query was that several Twitter users opine potentially sexist comments and signal so using the presence of this phrase, which arguably serves as a disclaimer for sexist opinions.\\ 

\textit{4.2 Was there any data that could not be adequately collected?}\\
For the parts of the dataset that stem from existing datasets used in the literature, only fractions of these datasets could be recovered, as many of their instances were deleted or removed from Twitter after having been posted and were thus unavailable for recollection.\\

\textit{4.3 Is any information missing from individual instances? Could there be a systematic bias?}\\ 
The keyphrase ``call me sexist, but" was removed from all the tweets collected for the dataset to avoid its priming effect on annotators, who have been shown to be more likely to consider a tweet sexist if it included the keyphrase. However, since this was done for all of the tweets in the dataset, there should be no systematic bias arising from this decision.\\ 

\textit{4.4 Does the dataset include sensitive or confidential information?}\\
The Twitter dataset was pseudonymized by replacing any mentions (@username) with a placeholder (MENTION). For the adversarial examples written by the MTurkers, their IDs were also pseudonymized. 
Mentions of family names, identified via regular expressions and a NER model, were manually confirmed to actually be of family names and then shortened to only the initial letter using a regular expression (e.g., John Doe to John D.).  
Even though the dataset might still contain tweets with sensitive or confidential contents, the described procedures prevent these contents from being easily associated with the corresponding individuals.\\

\textbf{User Selection Error}\\
The questions aiming at issues related to the ``User Selection Error" do not apply to this dataset, as the instances do not represent individuals. Furthermore, the collection and sampling of the data happened on the trace (tweet) level, not the user (Twitter user) level.\\ 

\textbf{\underline{5. Data Preprocessing and Data Augmentation}}\\

\textbf{Trace Augmentation and Trace Measurement Error}\\

\textit{5.1 Is there a label or target associated with each instance? If so, how were the labels or targets generated?}\\
Crowdworkers hired from Amazon Mechanical Turk were given the task of annotating the different instances for sexism. Each annotation consisted of one statement that crowdworkers were asked to annotate on two single-choice lists, containing the codes for sexist content and phrasing.
The toxicity scores associated with each instance were obtained from the automated toxicity classifier of the Perspective API.\\  

\textit{5.2 If automated methods were used, how does method performance impact the augmentations?}\\
The toxicity scores obtained from the Perspective API were derived automatically. The Perspective API runs on a ML model trained on comments from different online sources. This paradigm comes with a number of potential errors and biases that it might introduce through its annotations, as for example explained in this paragraph taken from the Perspective API’s model card:  
``Machine learning models learn from the data they’re trained with, so any biases in the data can creep into the predictions the models make. For example, our models sometimes predict higher toxicity scores for comments with terms for more frequently targeted groups (e.g. words like ``black", ``muslim", ``feminist", ``woman", or ``gay") because comments about those groups are over-represented in abusive and toxic comments in the training data."\\ 

\textit{5.3 If human annotations were used, who were the annotators that created the labels? How were they recruited or chosen? How were they instructed? How were they remunerated?}\\ 
Crowdworkers were recruited from Mechanical Turk and had to be located in the US, with over 10,000 HITs approved and over 99\% HIT approval rate. They furthermore had to pass a strict qualification test that ensured that they understood the construct of sexism as defined in the codebook, and did not apply overly subjective notions of sexism in their labeling. The test was passed if they correctly annotated 4 out of 5 ground-truth samples. This final test could only be taken once. 
The training imposed on annotators before taking the test described the codes for each of the two annotation categories, relying on examples and counter-examples. As additional guidance, the codebook developed based on the sexism scales was available to the annotators.
Crowdworkers received 6 cents per annotation, and 20 cents per counterfactual augmentation, resulting in a ``fair hourly wage".\\

\textit{5.4 If the final label was derived from multiple annotations, how was this done?}\\
Each instance was annotated by five annotators, and only those instances which at least three annotators considered as sexist, either because of content or phrasing, were marked as such. The final label was then obtained via majority vote.\\

\textit{5.5 Have there been any attempts to validate the labels?}\\
For the annotations generated by the crowdworkers, and using the majority rule for the five annotations provided for every sentence, Randolph’s Kappa for the content-annotations ($\kappa = .62$) and phrasing-annotations ($\kappa = .82$) were calculated and found to be satisfactory. The authors additionally report a majority agreement of 81\%, 98.8\% and 100\% for content, phrasing and overall sexism, respectively.\\

\textit{5.6 How could the data be misused?}\\
As the data contains texts that were identified as sexist and/or toxic, together with corresponding labels or scores, potential misuse scenarios of this data would include their utilization as a source for sexist content - be it by directly ``re-using" the instances of the dataset, or by using them as input or training data for systems that generate sexist content.\\

\textit{5.7 Could the dataset in any way contribute to the creation or reinforcement of social inequality?}\\ 
The dataset could potentially be used - together with other datasets of a similar type - to train or finetune a generative language model for the creation of sexist tweets, similar to those included in the dataset. Such a model could then be used generate and populate social media and web platforms with sexist content to harass vulnerable people, thereby reproducing and amplifying the harms and inequalities caused by the use of sexist language online.\\

\textbf{User Augmentation Error}\\

\textit{5.8 Have attributes and characteristics of individuals been inferred?}\\
There are no individuals explicitly identified in the dataset, thus there are neither observed nor inferred attributes and characteristics of individuals in the dataset.\\ 

\textit{5.9 Is it possible to identify individuals either directly or indirectly from the data?}\\
As there are no explicitly identified in the dataset, and as those individuals being referred to or mentioned in tweets have been pseudonymized, it should not be possible to neither directly nor indirectly identify individuals from the data.\\ 

\textbf{Trace Reduction Error}\\

\textit{5.10 Have traces been excluded? Why and by what criteria?}\\
No, all tweets collected via the previously described query have been retained.\\

\textbf{User Reduction Error}\\

\textit{5.11 Have users been excluded? Why and by what criteria?}\\
No, all Users corresponding to the tweets collected via the previously described query have been retained.\\

\textbf{Adjustment Error}\\

\textit{5.12 Does the dataset provide information to adjust the results to a target population? If so, is this information inferred or self-reported?}\\
The dataset does not contain any demographic information, and thus does not contain any information that would allow for any type of inference.\\

\subsection*{References}
\begin{enumerate}
    \item Buolamwini, J., \& Gebru, T. (2018). Gender shades: Intersectional accuracy disparities in commercial gender classification. In Conference on fairness, accountability and transparency (pp. 77-91).
    \item Samory, M., Sen, I., Kohne, J., Flock, F., \& Wagner, C. (2021). “Call me sexist, but...”: Revisiting Sexism Detection Using Psychological Scales and Adversarial Samples. In Proceedings of the International AAAI Conference on Web and Social Media (Vol. 15, pp. 573-584).
    \item Sen, I., Samory, M., Wagner, C., \& Augenstein, I. (2022). Counterfactually Augmented Data and Unintended Bias: The Case of Sexism and Hate Speech Detection. In Proceedings of the 2022 Conference of the North American Chapter of the Association for Computational Linguistics: Human Language Technologies (pp. 4716-4726).
    \item Wojcik, S., \& Hughes, A. (2019). Sizing Up Twitter Users. URL: https://www.pewresearch.org/internet/2019/04/24/sizing-up-twitter-users/ [last visited: 20.06.2032]
\end{enumerate}

\end{document}